# Nanoscale diffractive probing of strain dynamics in ultrafast transmission electron microscopy

Armin Feist†, Nara Rubiano da Silva†, Wenxi Liang‡, Claus Ropers†¶, Sascha Schäfer†*

† *4th Physical Institute - Solids and Nanostructures, University of Göttingen, Göttingen, Germany*
‡ *Wuhan National Laboratory for Optoelectronics, Huazhong University of Science and Technology, Wuhan, China*
¶ *International Center for Advanced Studies of Energy Conversion (ICASEC), University of Göttingen, Göttingen, Germany*

**ABSTRACT**

The control of optically driven high-frequency strain waves in nanostructured systems is an essential ingredient for the further development of nanophononics. However, broadly applicable experimental means to quantitatively map such structural distortion on their intrinsic ultrafast time and nanometer length scales are still lacking. Here, we introduce ultrafast convergent beam electron diffraction (U-CBED) with a nanoscale probe beam for the quantitative retrieval of the time-dependent local distortion tensor. We demonstrate its capabilities by investigating the ultrafast acoustic deformations close to the edge of a single-crystalline graphite membrane. Tracking the structural distortion with a 28-nm/700-fs spatio-temporal resolution, we observe an acoustic membrane breathing mode with spatially modulated amplitude, governed by the optical near field structure at the membrane edge. Furthermore, an in-plane polarized acoustic shock wave is launched at the membrane edge, which triggers secondary acoustic shear waves with a pronounced spatio-temporal dependency. The experimental findings are compared to numerical acoustic wave simulations in the continuous medium limit, highlighting the importance of microscopic dissipation mechanisms and ballistic transport channels.

Controlling confined phononic modes in the giga- to terahertz frequency range offers new approaches to steer the flow of heat in nanoscale structures[1] with a broad field of potential applications, ranging from advanced thermoelectric devices[2] to the heat management in dense semiconductor circuits.[3] Furthermore, coupled to tailored light fields, phononic modes with mega- to gigahertz resonance frequencies already developed into essential building blocks in nanometrology.[4,5]

Nanophononics based on tailored multilayer structures has made great progress in recent years, achieving, for example, phonon filtering,[6] and phonon amplification.[7] Beyond layered systems, three-dimensionally nanostructured materials facilitate thermally rectifying behavior,[8] highly efficient channeled thermal transport across nanoscale vacuum gaps,[9–11] enhanced light matter interactions in combined phononic-photonic resonators[12] and phonon lasing.[13,14] Optical methodologies, such as ultrafast optical spectroscopy[15] and Brillouin scattering,[16–18] allowed for experimental access to the spectral and temporal properties of nanophononic systems, including resonance frequencies, dissipation times[19] and nonlinear couplings.[20] However, extracting quantitative information on the structural distortion in nanophononic structures often requires elaborate theoretical modeling. Knowledge of the strain field is essential for tailoring the interaction between phononic fields and other degrees of freedom, such as the coupling of lattice distortions to the electronic[21] and magnetic[22,23] subsystems, interaction with confined light fields[12], and phase-transitions driven by acoustic[24] and optical[25] phonon fields.

In laterally homogenous samples, ultrafast electron[26–33] and X-ray[34–38] diffraction allows for quantitative access to collective transient lattice distortions. Extending these approaches to three-dimensionally nanostructured geometries remains challenging, despite recent progress in ultrafast coherent diffractive dark-field imaging[39,40] utilizing intense X-ray pulses at free-electron laser facilities.[41] In a table-top approach, ultrafast transmission electron microscopy (UTEM)[42–50] provides a visualization of nanophononic modes by time-resolved bright-field imaging,[51–54] with first steps towards local diffractive probing.[55–57] However, the full capabilities of conventional transmission electron microscopy[58–64] for the quantitative mapping of strain fields has not been harnessed in UTEM.

Here, we demonstrate the quantitative nanoscale probing of optically triggered ultrafast strain dynamics in UTEM, employing ultrashort electron pulses in convergent beam electron diffraction (CBED). We achieve a quantitative three-dimensional spatio-temporal reconstruction of the ultrafast lattice distortions in nanoscopic volumes close to the edge of a single crystalline graphite membrane. High-amplitude coherent expansional and shear acoustic waves are launched at the symmetry-breaking sample boundaries, and we track their ballistic propagation and dephasing on nanometer length scales.

In the experiments, we generate low-emittance ultrashort electron pulses by localized photoemission from a tip-shaped field emitter[45,50,65] (see Supporting Information, SI 1). The femtosecond electron pulses are accelerated to an electron energy of 120 keV, coupled into the electron optics of a transmission electron microscope and tightly focused (28-nm focal spot size) onto a 120-nm thick graphite membrane (see Supporting Information, SI 2). For varying probing positions relative to the edge of the membrane, electron diffraction patterns are detected in the far-field (Figure 1a). The sample is optically excited by femtosecond laser pulses (800-nm central wavelength, 50-fs pulse duration, 50-μm focal spot diameter, 16-mJ/cm² fluence), and, at an adjustable delay time $\Delta t$ relative to the electron pulse arrival, local structural dynamics are stroboscopically mapped at the electron focal spot position.

Figure 1c displays a typical ultrafast large-angle convergent beam electron diffraction pattern recorded with femtosecond electron pulses before optical excitation ($\Delta t < 0$). In the pattern, the central intense disc-like feature represents the angular distribution of the illuminating electron pulses.

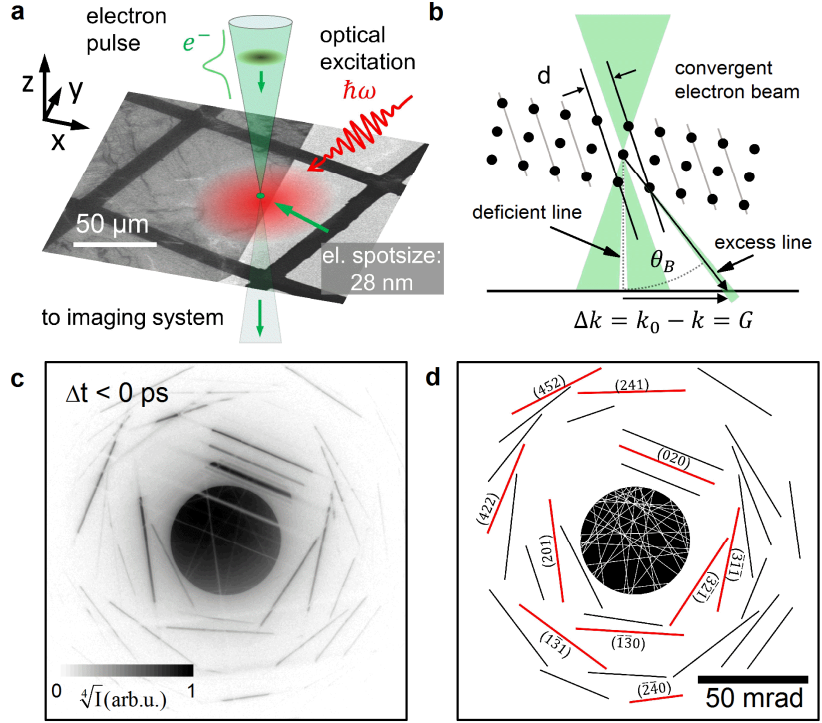

Figure 1. Ultrafast convergent beam electron diffraction on single crystalline graphite. (a) Local diffractive probing (28-nm electron focal spot size) of optically induced inhomogeneous structural dynamics in a single crystalline graphite membrane (background: overview bright field electron micrograph). (b) Bragg scattering from lattice planes (*hkl*) results in a momentum transfer $\Delta \mathbf{k} = \mathbf{G}_{hkl}$, forming deficit and excess lines in the diffraction pattern. (c) CBED pattern before optical excitation (exemplary probing position: 500-nm distance to crystal edge). For better visibility of diffraction lines at high scattering angles, the fourth root of the electron intensity *I* is shown. (d) Calculated deficit (white) and excess (black, red) Bragg line positions for the employed sample orientation. For selected Bragg lines, the corresponding Miller indices are given.

Bragg scattering conditions for the graphite lattice planes (*hkl*) are fulfilled along specific lines in momentum space[66]. At their intersection with the central disc, efficient scattering occurs, forming deficit intensity lines within the disc, and excess lines, which are radially displaced by the Bragg angle $\theta_B$ (Figure 1b,d).

The angular displacement of each line encodes the length and orientation of a specific reciprocal lattice vector $\mathbf{G}_{hkl}$ and the corrugation of the scattering potential.[61,67] Thereby, U-CBED gives access to the ultrafast temporal change of local lattice periodicities $d_{hkl}$ and atomic mean-square displacements $\sqrt{\langle u^2 \rangle}$ (see Supporting Information, SI 3). The broad angular range of the incident electron beam (50 mrad full convergence angle) and the chosen sample orientation enable the simultaneous observation of multiple independent Bragg scattering conditions and the corresponding rocking curves,[68] providing for direct experimental access to the local structural distortion and its temporal evolution.

After optical excitation, we observe pronounced delay-dependent radial Bragg line shifts $\Delta \theta$ (by up to 6 mrad) in the CBED pattern (for a series of delay-dependent diffraction pattern, see movie Supporting Information, M1 and M2). The induced strain dynamics results in no significant azimuthal rotation of Bragg lines for the chosen sample orientation. In the following, we therefore consider the transient changes of Bragg line profiles, obtained by integrating the diffracted intensity along the individual line directions.

In Figure 2, we show the delay-dependent profiles of selected excess Bragg lines for two different probing positions. With the electron focal spot placed at a distance of $r = 500$ nm from the edge of the graphite membrane (Figure 2b), the (422) and ($\bar{3}\bar{2}\bar{1}$) lines display a strong multi-frequency oscillatory behavior of the average line position and a modulation of the line profile, even including line splittings into multiple components. Other Bragg lines show a different temporal characteristic (e.g. ($\bar{2}\bar{4}0$)), or only very weak overall changes (e.g. (020)). Remarkably, the recorded transient changes are strongly influenced by the nearby membrane edge. In a continuous part of the film, a much simpler dynamic behavior of the line profiles is observed, as is evident by comparing the transient (422) profiles in Figures 2a and b.

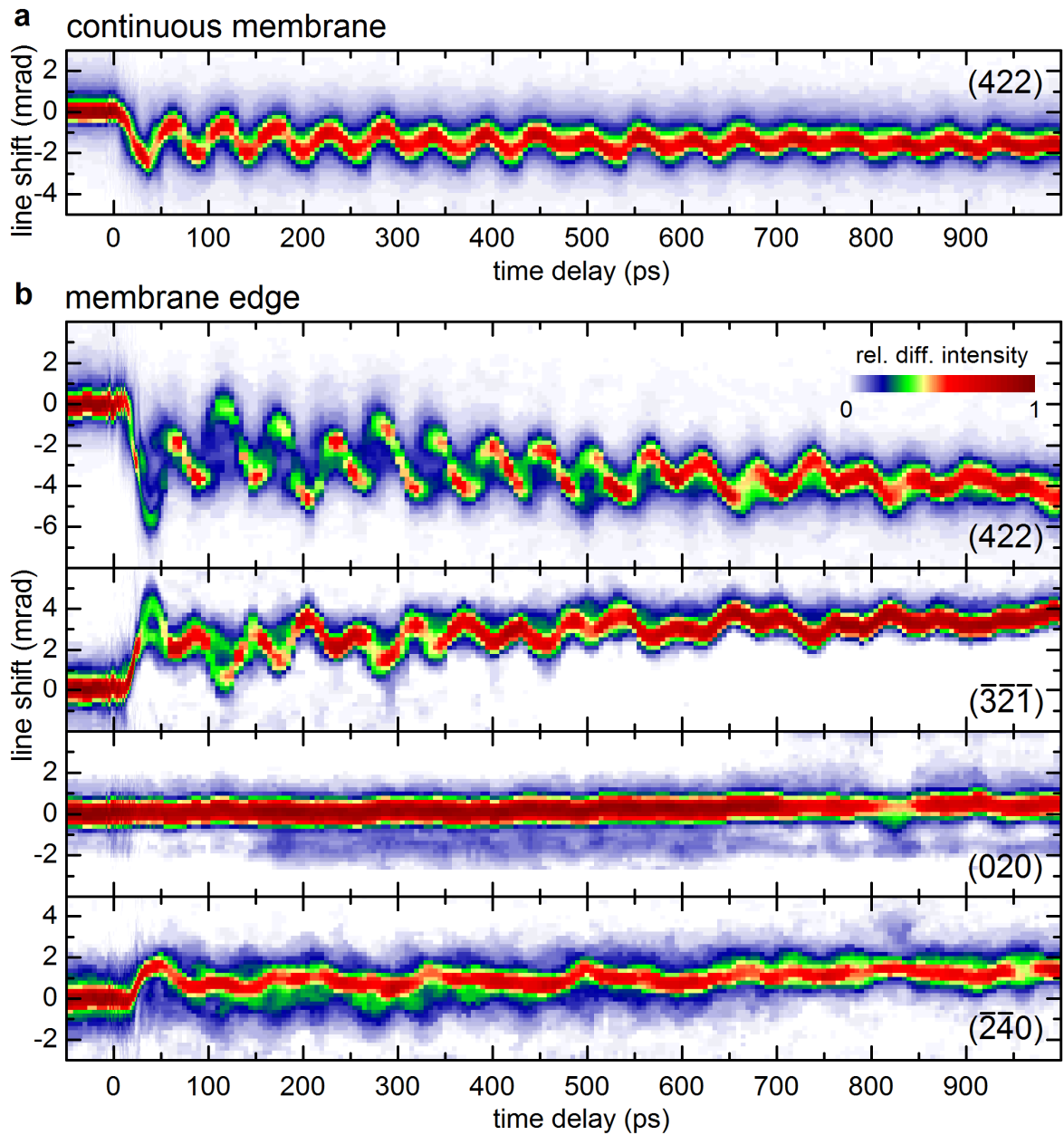

Figure 2. Transient modulation of Bragg line profiles. Delay-dependent profiles of selected Bragg lines for probing (a) within a continuous part of the membrane, and (b) close to the graphite membrane edge (500-nm relative distance).

A quantitative analysis of the Bragg line shifts in Figure 2 allows us to identify the complex superposition of the acoustic lattice distortions involved in the optically driven dynamics at a homogeneous part of the membrane and at its edge. Microscopically, the evolution of the local structural deformation of the graphite film is described by a time-dependent distortion tensor field $\mathbf{F}(r, \Delta t) = \boldsymbol{\varepsilon} + \boldsymbol{\omega} + \mathbf{I}_3$, which can be decomposed into a symmetric strain tensor $\boldsymbol{\varepsilon}$ and an antisymmetric rotation tensor $\boldsymbol{\omega}$ ($\mathbf{I}_3$: unit tensor). The local distortion alters the spacing and orientation of crystal lattice planes, resulting in characteristic shifts of Bragg conditions in momentum space (Figure 3a). We extract local distortion tensor components for each delay time considering the center-of-mass of multiple experimental Bragg line positions (selected traces shown in Figure 3c, see Supporting Information, SI 3).

The temporal evolution of the distortion tensor near the membrane edge is dominated by two components $F_{zz}(\Delta t) = 1 + \varepsilon_{zz}$ and $F_{xz}(\Delta t) = (\varepsilon+\omega)_{xz}$ (Figure 3d,e, red curves), corresponding to an expansional strain along the graphite out-of-plane $z$-axis (for coordinate system see Figure 1a and Supporting Information Figure S3) and a shear-rotation in the $xz$-plane (perpendicular to the membrane edge), respectively. Both deformations leave the (0k0) lattice planes unchanged, consistent with the experimentally found negligible transient changes of the (020) line profiles (cf. Figure 2b). Remarkably, the distortion tensor analysis disentangles the multi-frequency temporal behavior of individual Bragg line shifts. The components $\varepsilon_{zz}$ and $(\varepsilon+\omega)_{xz}$ each exhibit damped oscillations at a single frequency, with periods of $T_{expansion} = 56.5 \pm 1.6$ ps and $T_{shear-rot} = 154 \pm 5$ ps (central frequencies of $17.7 \pm 0.5$ and $6.5 \pm 0.2$ GHz, see Figures 3d and 3e) for the expansional and shear motion, respectively. Far from the membrane edge (150-µm distance), the optically induced deformation (Figure 3d,e, black curves) is primarily governed by the expansional out-of-plane motion, and no significant amplitude in the $xz$-component of the distortion tensor is observed.[69]

The periods of the expansional and shear-rotational distortions, $T_{expansion}$ and $T_{shear-rot}$, are given by the roundtrip time of the acoustic waves propagating between the two faces of the membrane. The ratio $T_{shear-rot}/T_{expansion} = 2.73 \pm 0.16$ is in excellent agreement with the relative magnitude of the corresponding longitudinal and transverse acoustic sound velocities in single crystalline graphite $v(LA[001])/v(TA[001]) = (4140 \text{ m/s})/(1480 \text{ m/s}) = 2.80$.[70] Furthermore, the periods $T = 2l/v$ yield a membrane thickness of $l = 117$ nm, which matches the value of 120 nm derived by evaluating thickness fringes[66,71] in the CBED patterns.

At long delay times ($\Delta t > 800$ ps), the oscillatory membrane expansion becomes strongly damped, approaching an average graphite interlayer distance increase of about 1.5% at the membrane edge (continuous membrane: 0.9%). In order to compare these strain values to a thermal expansion model, we extract the local temperatures from the integrated diffracted intensity change of the (452) Bragg-line after optical excitation (Figure 3b). For an equilibrated phonon distribution ($\Delta t > 100$ ps), a thermal Debye-Waller behavior is reached and we extract an optically induced temperature rise of $\Delta T_{cont} = 270$ K at a continuous part of the membrane and $\Delta T_{edge} = 480$ K at the membrane edge, which corresponds to thermal film expansions of 0.93% and 1.65%, respectively (see Supporting Information, SI 5). Importantly, ultrafast CBED directly yields full transient rocking curves, so that an acoustic lattice distortion (line shift) and a change in the atomic mean square displacement (line intensity) can be distinguished.

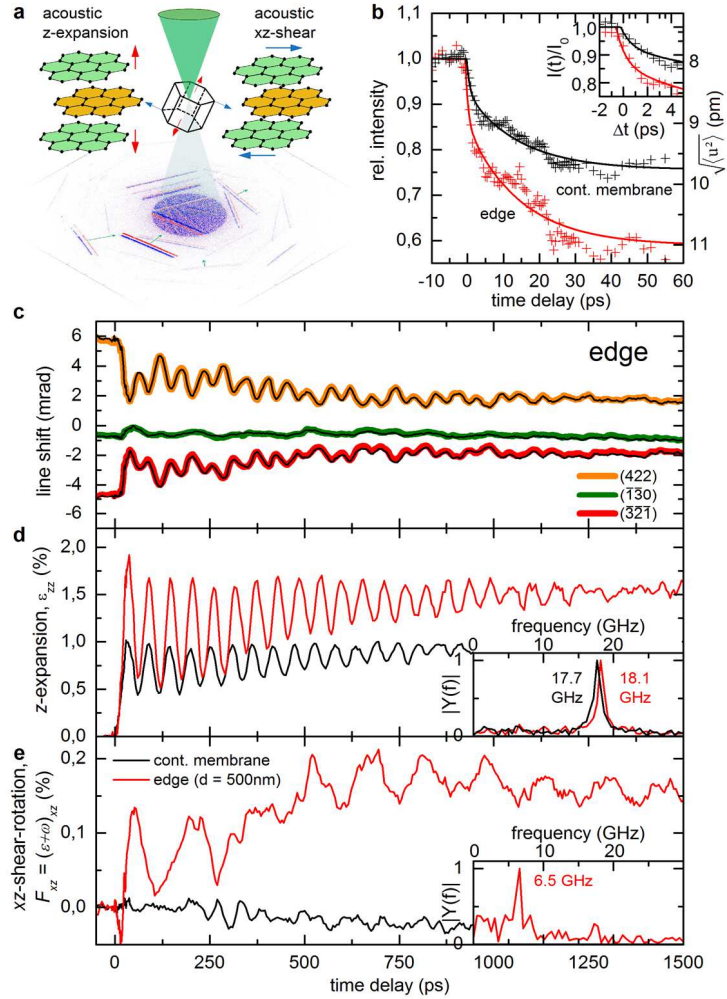

Figure 3. Time-dependent Bragg-line changes and dynamics of distortion tensor components. (a) Local probing of the mean unit cell deformations reveals two dominating mechanical modes: an out-of-plane $z$-axis expansion and an acoustic shear-rotation in the $xz$-plane. (b) Change in (452) Bragg-line intensity and square root of atomic mean square displacement $\sqrt{\langle u^2 \rangle}$ in the in-plane direction after optical excitation for probing at the graphite edge (red) and in a continuous part of the membrane (black). (c) Experimentally obtained delay-dependent center-of-mass shift (black line) and reconstructed mean line position (colored line, background) of the (422), ($\bar{1}\bar{3}0$) and ($\bar{3}\bar{2}\bar{1}$) Bragg lines, probed at the graphite edge. (d,e) Reconstructed $z$-axis expansion (d) and in-plane $xz$-shear-rotation (e) components (red: edge, black: cont. membrane) with respective Fourier analysis (inset, |Y(f)|: Fourier amplitude).

At early delay times, a biexponential drop of diffracted intensity is observed, which is attributed to the previously reported initial non-thermal phonon distribution after optical excitation.[72–74] This delayed increase in atomic mean square displacement is also reflected in a phase shift of the out-of-plane breathing oscillation. Specifically, we observe the first maximum of $\varepsilon_{zz}$ at 36 ps, corresponding to a considerable time lag of about 7 ps relative to a cosine-like transient. The quantitative relation between the non-equilibrium atomic mean square displacement and the resulting stress in the in-plane and out-of-plane directions requires further study, potentially contributing to elucidate the complex hierarchy of energy dissipation in graphite.[30,72–78]

The out-of-plane expansional breathing modes, visible in $\varepsilon_{zz}$, are universal features observed in laser excited thin films as a result of a transient stress gradient $\sigma(z)$ within in the depth of the film, with electronic and lattice contributions.[15,30,79–83] For the generation of shear modes, as mapped in $F_{xz}$, a symmetry breaking in the lateral direction is required, such as in anisotropic or strained crystal lattices or by local light fields.[84–88]

In our sample geometry, the structural symmetry is locally broken on mesoscopic length scales due to the presence of the membrane edge. Ultrafast CBED now allows for a local mapping of the evolving distortion tensor field and the sources of the corresponding acoustic waves. To this end, we record time-resolved local diffraction patterns with the focused electron pulses placed at varying distances $r$ from the membrane edge. Figure 4a exemplarily shows the angular shift of the (201) Bragg line as a function of the delay time $\Delta t$ and the probing position $r$, together with the extracted distortion tensor components $F_{zz}(r, \Delta t) = 1 + \varepsilon_{zz}(r, \Delta t)$ and $F_{xz}(r, \Delta t)$ (Figures 3b,e).

The expansional mode is observed at all probing positions with an equal phase. Its amplitude is spatially modulated and in particular at $r = 500$ nm is increased by about 70% compared to the value found at a larger distance from the graphite edge. This ratio agrees well with the larger temperature rise at this probing position, as observed by the transient Debye-Waller behavior (see Figure 3b). The locally increased sample excitation can be attributed to an interference pattern formed by the optical excitation close to the membrane edge, which is also observable in optically driven inelastic electron scattering, utilizing scanning photon-induced near-field electron microscopy (S-PINEM)[45,89–91] (Figure 4d, see Supporting Information, SI 7).

In contrast to the film breathing mode, the shear-rotation component of the distortion tensor $F_{xz}$ shown in Figure 3e exhibits a pronounced spatial dependence. In particular, the onset time of $F_{xz}$ scales linearly with the distance from the membrane edge, with a slope corresponding to a phase velocity of ~22 km/s.

To further analyze the peculiar spatio-temporal strain dynamics, we numerically solve the elastodynamic wave equation for our sample geometry, considering a thermal stress model and a laterally homogeneous sample excitation profile (see Supporting Information, SI 6). For the $\varepsilon_{zz}(r, \Delta t)$ component, we obtain an $r$-independent temporal evolution (Figure 4c (top)), in agreement with the breathing mode of a continuous membrane. In addition, optical excitation results in an in-plane thermal stress $\sigma_x$ of the graphite lattice, which launches an expansional shock wave in $\varepsilon_{xx}(r, \Delta t)$ from the membrane edge (Figure 4c (bottom)), propagating perpendicular to the edge with the longitudinal in-plane sound velocity LA[100] = 22.16 km/s.[70] Due to the optical excitation profile, the expansional in-plane shock wave is localized to the top of the membrane and thereby induces shearing of the thin film sample. The ultrafast build-up of shear strain at the top initiates the shear wave travelling back and forth between the membrane faces.[92] This model readily explains the local excitation of the experimentally observed shear wave with its onset time scaling linear with the distance to the graphite edge. In particular, for a pure film shearing along the [100] direction the corresponding strain and rotation components $\varepsilon_{xz}$ and $\omega_{xz}$ are of equal magnitude, so that $F_{xz}/2 = \varepsilon_{xz} = \omega_{xz}$ and $F_{zx} = 0$.

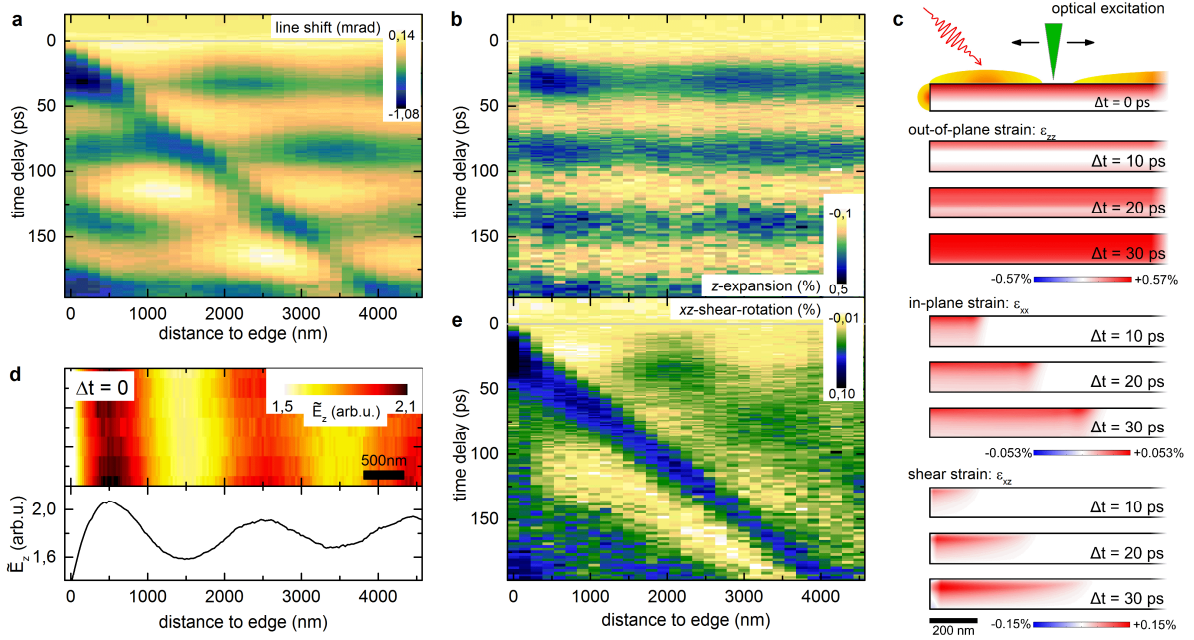

Figure 4. Spatio-temporal mapping of near-edge strain dynamics in single crystalline graphite. (a) Mean shift of the (201) Bragg-line as a function of time delay and probing position (recorded at a reduced optical fluence compared to the non-spatially resolved CBED data). Extracted z-axis expansion (b) and in-plane shear-rotation (e) retrieved by evaluating the shift of several Bragg lines. (c) Numerical finite-element simulation of the $\varepsilon_{zz}(r, \Delta t)$, $\varepsilon_{xx}(r, \Delta t)$ and $\varepsilon_{xz}(r, \Delta t)$ strain tensor components (absorbed energy density adopted to match experimental $\varepsilon_{zz}$ strain amplitude), illustrating the out-of-plane expansion and the in-plane propagating shock wave within 30 ps after optical excitation. (d) Characterization of the optical near-field structure by scanning photon-induced near-field electron microscopy (S-PINEM), with optical incidence angle of about 39°.

Up to here, Bragg line shifts in scanning U-CBED yielded a spatio-temporal map of the lateral structural distortion of the photo-excited graphite membrane. In addition, rich experimental information on the inhomogeneous strain within the depth of the membrane is contained in the profiles of the Bragg lines, which we analyze in the following. Within kinematic scattering theory[93], a strained crystal imprints a phase modulation onto the diffracted electron wavefront,[39,58] resulting in a CBED profile well described by

$$I(\Delta\theta \cdot |\mathbf{G}_{hkl}|) \propto \left|\mathcal{F}\left(e^{i\mathbf{G}_{hkl}\cdot\mathbf{u}(z)}\right)\right|^2 \quad \text{(Eq. 1)}$$

in which $\Delta\theta$ is the change in diffraction angle (relative to the Bragg angle $\theta_B$), $\mathbf{G}_{hkl}$ the corresponding reciprocal lattice vector, $\mathbf{u}(z)$ the atomic displacement field, and $\mathcal{F}$ the Fourier transformation along the graphite z-axis (see Supplement). The corresponding distortion tensor $\mathbf{F}$ is given (for small deformations, as relevant here) by the gradient of the displacement field, i.e. $\mathbf{F} = \mathbf{I}_3 + \nabla\mathbf{u}$. Notably, the line profiles depend on the projection $\mathbf{G}_{hkl}\cdot\mathbf{u}(z)$ (Eq. (1)), so that the cross sections for individual Bragg conditions are sensitive to different components of the displacement vector field and thereby to the polarization of the involved phonon modes.

In Figure 5, we exemplarily compare the experimental time-dependent (422) line profiles at the membrane edge and in the continuous film with predicted profiles according to Eq. 1, utilizing the numerically simulated displacement fields. For the continuous part of the graphite film, a periodic change of the Bragg line width is observed (with a period $T_{\text{expansion}}/2$), which is well reproduced within the numerical strain model (Figure 5a and 5c (left panel)). Approximately at delay times of maximum film expansion and compression, sharp Bragg lines are obtained due to the intermediate nearly homogeneous $\varepsilon_{zz}$ strain distribution within the film, as for example visible in Fig 4c at $\Delta t$ = 30 ps. The slight time lag between Bragg line shift and line broadening as well as their relative amplitude, sensitively depend on the optical excitation depth and the resulting transient stress profile. In particular, the experimental width of the Bragg line profiles cannot be reproduced if one considers the optical penetration depth in graphite of $\delta_P$ = 36 nm[94] alone. Instead, a good agreement is obtained for an excitation depth spatially spread to about 90 nm (see Supplement), which may be caused by fast interlayer electron or ballistic phonon transport.[95,96] Furthermore, the asymmetry at the crests of the oscillatory Bragg line movement is reproduced well in the simulations by adopting a 8-ps coupling time of the initial excitation to the experimentally detected coherent out-of-plane motion, similar to the time constant observed for the increase of the in-plane atomic mean square displacement.[72–74]

For the strain dynamics induced at the membrane edge, the more complex behavior of Bragg line profiles (Figure 5b and 5c (right panel)) is a result of the superposition of expansion and shear deformation, resulting in different projections of the displacement field $\mathbf{u}(z)$ onto reciprocal lattice vectors $\mathbf{G}_{hkl}$ (cf. Eq. 1). The main features of the experimental line shapes are regained in the numerical strain simulation, including the decreasing intensity maximum after $\Delta t$ = 0 with a pronounced line sub-structure between 23 and 60 ps (Figure 5b). In addition, also the general experimental trend of partial re-focusing between 60 and 90 ps, and increased line broadening between 90 and 140 ps is found in the simulation. Microscopically, the line shapes sensitively depend on the relative amplitudes and phases of the expansional and shear wave modes, allowing for a sensitive mapping of nanophononic strain fields. Remaining difference between the experimental and simulated line profiles may indicate the breakdown of classical continuum mechanics at the length and time scales considered here. Further developments are required to properly account for the initial non-thermal phonon distribution and mode specific phonon-phonon interactions, and their impact on ultrafast transport processes and the transient local lattice stress, particularly relevant for the nanoscale geometries considered here.

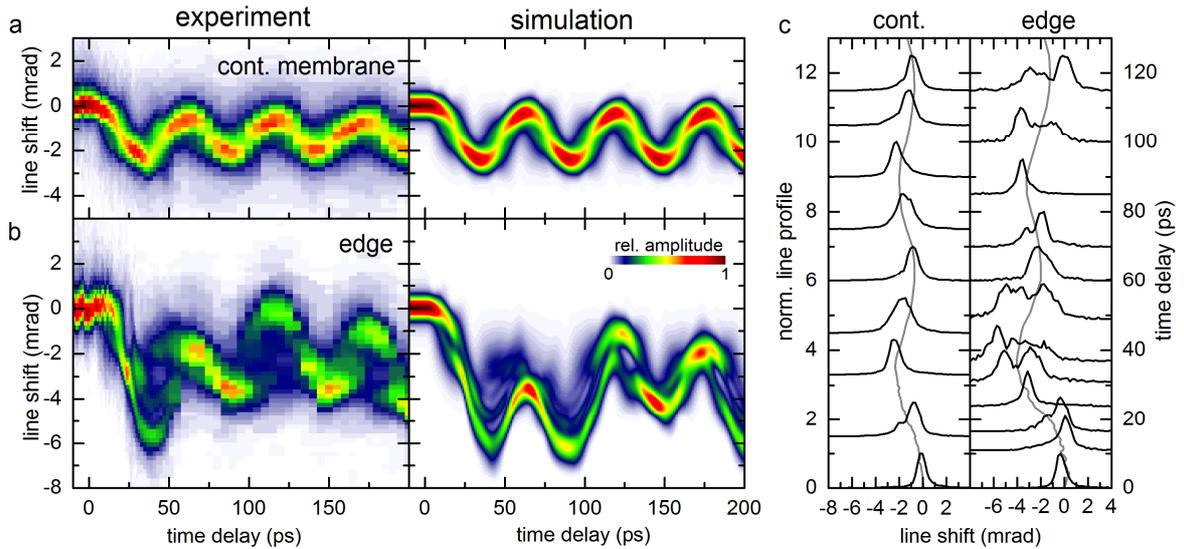

Figure 5. Dynamics of the (422) Bragg line profile. Extracted time-dependent cross-sections of the (422) line in (a) a continuous part of the membrane and (b) close to the graphite edge (500 nm distance) compared to calculated line profiles with displacement fields derived from numerical simulations. (c) Normalized line profiles (left-hand axis) at specific time delays (right-hand axis), with mean line position indicated in gray.

In conclusion, we demonstrated the quantitative mapping of a time-dependent distortion tensor field in a nanoscale geometry, utilizing ultrafast convergent beam diffraction with a raster-scanned ultrashort electron probe. Our technique is applicable to a wide variety of locally structured thin-film sample systems. In particular, we believe that U-CBED opens a new avenue for achieving a quantitative description of ultrafast processes relevant in nanophononic devices, potentially allowing for a precise tailoring of nanostructure and function. With the temporal resolution demonstrated here, U-CBED is also capable to image phonon modes up to the terahertz regime, which will enable to address the flow of thermal energy on its intrinsic time and length scales. Such capabilities may help to unravel the influence of local dissipation channels in complex materials, transport processes across designed interfaces and nonlinear phononic interactions.

## ASSOCIATED CONTENT

The following files are available free of charge.

Supporting Information with details on the experimental setup (SI 1), sample preparation (SI 2), data collection & analysis (SI 3), distortion tensor extraction from the CBED pattern (SI 4), Debye Waller analysis (SI 5), numerical simulations (SI 6) and near field characterization (SI 7) (PDF).

Movie showing delay-dependent change in CBED intensity (difference pattern) probed in a continuous part of the membrane (M 1) and close to its edge (M 2) (AVI).

## AUTHOR INFORMATION

Corresponding Author

*E-mail: sascha.schaefer@phys.uni-goettingen.de

ORCID

Armin Feist: 0000-0003-1434-8895

Sascha Schäfer: 0000-0003-1908-8316

## AUTHOR CONTRIBUTIONS

A.F. conducted the time-resolved diffraction experiments and analyzed the data with contributions from W. L. and S.S.. N. R. prepared the single crystalline graphite membrane. Finite-element strain modelling was performed by A.F. and S.S. The manuscript was written by A.F. and S.S. with contributions from all authors. C.R. and S.S. conceived and directed the study. All authors discussed the results and the interpretation.

## ACKNOWLEDGMENT

We gratefully acknowledge funding by the Deutsche Forschungsgemeinschaft (DFG-SPP-1840 'Quantum Dynamics in Tailored Intense Fields', and DFG-SFB-1073 'Atomic Scale Control of Energy Conversion', project A05), support by the Lower Saxony Ministry of Science and Culture and funding of the instrumentation by the DFG and VolkswagenStiftung. W.L. would like to acknowledge support by the Director Fund of WNLO (Grant WNLOZZYJ1501) and the National Natural Science Foundation of China (Grant 11574094).

# Supporting Information for "Nanoscale diffractive probing of strain dynamics in ultrafast transmission electron microscopy"

Armin Feist†, Nara Rubiano da Silva†, Wenxi Liang‡, Claus Ropers†¶, and Sascha Schäfer†*

*† 4th Physical Institute - Solids and Nanostructures, University of Göttingen, Göttingen, Germany*
*‡ Wuhan National Laboratory for Optoelectronics, Huazhong University of Science and Technology, Wuhan, China*
*¶ International Center for Advanced Studies of Energy Conversion (ICASEC), University of Göttingen, Göttingen, Germany*

**SI 1: Ultrafast Convergent Beam Electron Diffraction (U-CBED)**

In ultrafast transmission electron microscopy (UTEM), the transient state of an optically excited sample is probed by ultrashort electron pulses. The broad imaging, diffraction and spectroscopy capabilities of state-of-the-art TEM, give access to ultrafast dynamics in different degrees-of-freedoms.[1–3]

The Göttingen UTEM instrument is based on a field emission TEM (JEOL JEM-2100F), which we modified to allow for optical sample excitation and femtosecond electron pulse generation from a laser-driven photocathode. In our photoelectron gun approach, we make use of localized single-photon photoemission from the apex of a Schottky-type ZrO/W field-emission tip at an optical wavelength of 400 nm, yielding low-emittance electron pulses (focus diameter down to 9 Å, pulse duration down to 200 fs (full-width-at-half-maximum (FWHM)), energy width of 0.6 eV in the space-charge free regime).[3]

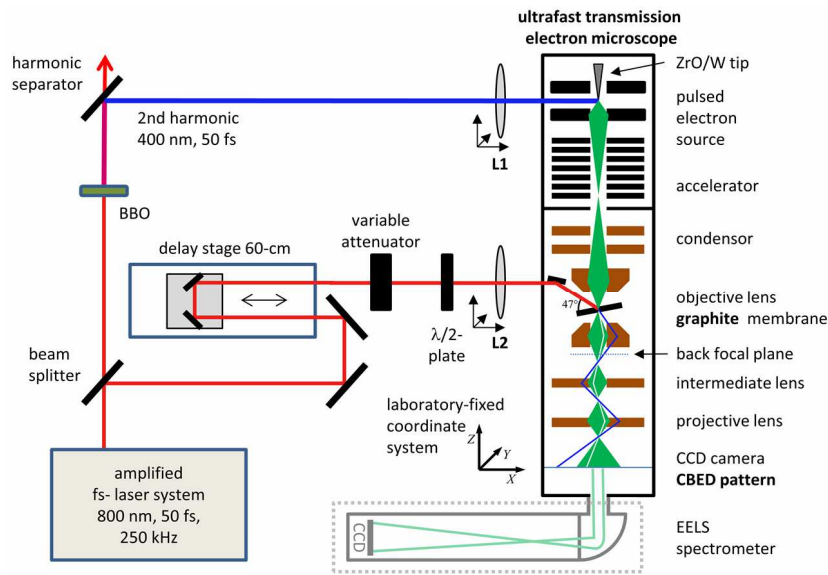

**Figure S1: Experimental Setup.** Femtosecond optical pulses from a Ti:sapphire amplified laser system (800 nm central wavelength, 50 fs pulse duration, 250 kHz repetition rate) are split in two optical paths. The first beam is frequency-doubled and focused (4.1 nJ pulse energy, ~0.3 mJ/cm², 50 fs pulse duration) on the nanoscopic tip emitter for the generation of ultrashort electron pulses (probe pulses). The second beam is attenuated and focused (320 nJ pulse energy, 16.2 mJ/cm² fluence, 50 µm focal spot size, p-polarized, 47° angle of incidence relative to the sample) with a well-defined relative timing set by a mechanical delay stage (see Ref. 3 for more details). The projection system of the TEM is used to record images in real and reciprocal space, as well as electron energy loss spectra with an additional spectrometer.

In the current experiment, the photoelectron gun was operated in a high transmission mode[3] at a 250-kHz repetition rate and an acceleration potential of 120 kV. At the sample position, the electron beam was focused with a full convergence angle of 50 mrad (200-µm condenser aperture) to a focal spot of 28 nm in diameter (spot size limited by spherical aberration of the objective lens, spherical aberration coefficient $C_s$ = 1.4 mm, focal length f = 2.7 mm), exhibiting a pulse duration of 700 fs (FWHM). The probing position on the sample can be varied with nanometer accuracy using the magnetic deflector coils above the objective lens. For optical excitation, p-polarized 50-fs laser pulses were focused onto the sample at an angle of 47° with respect to the membrane surface. Convergent beam diffraction patterns were recorded on a calibrated charge-coupled device (Gatan Ultrascan 4000), binned to 1024x1024 pixels (0.062 1/nm per pixel momentum resolution) with an effective projection camera length of 28.7 cm.

**SI 2: Sample**

The single crystalline membrane was prepared by mechanically cleaving a graphite mineral crystal (vendor: Naturally Graphite), resulting in 100-µm scale sample regions with homogeneous thickness. The graphite flake was supported on a copper grid (square 200 mesh, open areas of 114 x 114 µm$^2$).

For the probing position close to the graphite edge, we measure a membrane thickness of 120 nm by evaluating the CBED interference fringes recorded with a continuous electron beam.[4,5] Near-edge structural dynamics were acquired at a part of the membrane, in which the cleaving process produced a straight edge oriented along the [1$\bar{2}$0]-direction (cf. Figure S3), i.e. in an armchair-type configuration.[6]

For the time-resolved CBED experiments, the specimen was rotated by about 8° around the graphite y-axis and 1° around its x-axis (sample coordinate system, cf. Figure S3a). Hereby, a set of Bragg lines (*hkl*) is obtained, which yields a high sensitivity to acoustic waves with in-plane polarization parallel and perpendicular to the graphite edge, as well as out-of-plane polarizations. Precise angular alignment was regularly checked by comparing Bragg line positions to numerically calculated diffraction patterns (see below).

**SI 3: Data collection & analysis**

All diffractograms were integrated for 5 s, typically containing about 2.5 - 3.5 · 10$^6$ electrons. The delay time scans (Figures 1-3 and 5) consisted of 489 steps (75 min measurement time each) and were repeated multiple times (continuous membrane: 10 repetitions, close to the edge: 3 repetitions). For the spatio-temporal map at the graphite edge (Figure 4) the electron focal spot was shifted perpendicular to the graphite edge (41 spatial probing positions) for each individual delay time (159 time steps), resulting in a measurement time of about 16 h.

After normalization to the total number of electrons in each image and flat-field correction for residual aberrations, rectangular regions of interest (ROI) are defined, with one edge parallel to the respective Bragg line (*hkl*) (cf. Figure S2a). For each time step, a line profile is extracted (Figure S2b) by rotating the image parallel to the respective line and integrating the intensity within the ROI along the line direction. For each ROI, a delay-time averaged quadratic background is subtracted (Figure S2c,d). Amplitude, position and width of the Bragg lines are either extracted by a Lorentzian fit (Figure S2e, applied for weakly broadened line profiles, e.g. (020), and for all lines at low optical excitation density), or by calculating the center of mass and standard deviation (non-Lorentzian line shapes, e.g. (422) for a high excitation density).

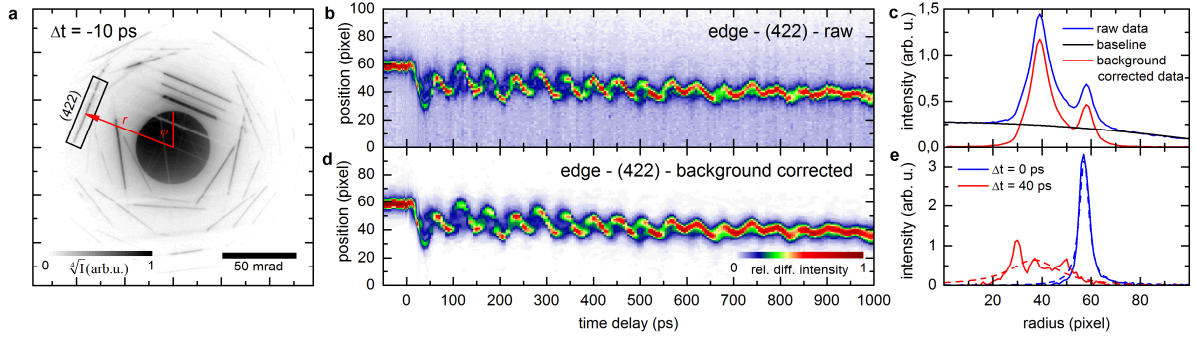

**Figure S2: Evaluation of the diffraction data.** (a) Raw image with indicated region of interest for a specific Bragg line. (b) Extracted line profile from raw data as function of time delay. (c) Removal of global background from delay-time integrated intensity of (b) (after normalization to the total number of electron per image). (d) Time dependent line-profiles after background removal and median filtering (along *r*-axis). (e) Exemplary line profiles (solid) with Lorentzian fit (dashed). Non-broadened profiles (e.g. $\Delta t = 0$ ps, blue) can be approximated by a Lorentzian function, while center-of-mass and standard deviation analysis are crucial for inhomogeneously broadened lines (e.g. $\Delta t = 40$ ps, red)

## SI 4: Extracting the lattice distortion tensor from CBED patterns

The quantitative analysis of CBED pattern requires an adequate description of the Bragg scattering conditions in k-space, which we derive from the graphite unit cell[7] defined by

$$\mathbf{a}_1 = a[1, 0, 0], \mathbf{a}_2 = a\left[1/2, \sqrt{3}/2, 0\right], \mathbf{a}_3 = c[0, 0, 1],$$

with lattice constants $a = 2.46$ Å and $c = 6.71$ Å. To account for an arbitrary sample orientation, the real space basis, represented by the matrix $\mathbf{B}_u = [\mathbf{a}_1, \mathbf{a}_2, \mathbf{a}_3]$ in the three-dimensional coordinate system (*x*, *y*, *z*), is rotated (see Figure S3a) by applying a matrix

$$\mathbf{R} = \mathbf{R}_Z(\gamma) \cdot \mathbf{R}_Y(\beta) \cdot \mathbf{R}_X(\alpha),$$

with rotation matrices $\mathbf{R}_{X,Y,Z}$ around the laboratory-fixed *X*, *Y* and *Z* axis, respectively. Taking the planar sample orientation into account, the angles $\alpha$ and $\beta$ correspond to the angular degrees-of-freedom of the double-tilt sample holder, and $\gamma$ is related to the azimuthal orientation of the graphite flake (cf. Figures 1 and S3). The reciprocal basis in the laboratory-fixed coordinate system is given by $\mathbf{G}_u = (\mathbf{R} \cdot \mathbf{B}_u)^{-1}$, so that the reciprocal lattice vector with Miller indices *h*, *k* and *l* is expressed as $\mathbf{G} = [h, k, l]^T \cdot \mathbf{G}_u$. For scattered and incident wave vectors $\mathbf{k}$ and $\mathbf{k}_0$, allowed scattering condition are obtained from

$$\mathbf{G}^2 + 2\mathbf{k}_0\mathbf{G} = 0$$

by considering the Laue equation $\mathbf{G} = \Delta \mathbf{k} = \mathbf{k} - \mathbf{k}_0$ (conservation of momentum) and elastic scattering $(\mathbf{G} + \mathbf{k}_0)^2 = \mathbf{k}^2$ (conservation of energy).[8]

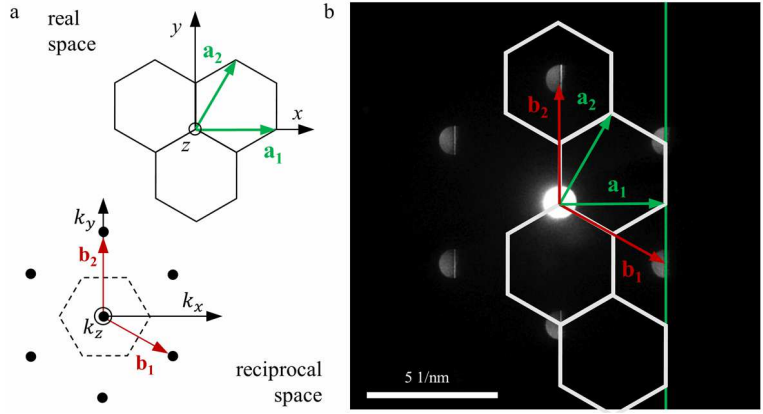

**Figure S3: Sample orientation.** (a) In-plane graphite unit cell in real and momentum space with corresponding real space coordinate system. (b) Parallel illumination, strongly defocused diffraction pattern with half of the beam clipped by graphite edge shows the relative orientation of real and k-space in the diffractograms. The overlaid real space hexagonal unit cells illustrates the armchair configuration of the edge.

In the paraxial approximation, i.e. $k_{0X}, k_{0Y} \ll k_0$ and $k_X, k_Y \ll k$ (for the optical axis chosen along $Z$), the scattering conditions can be simplified to

$$-G^2/2 = k_X G_X + k_Y G_Y + k_0 G_Z$$

which describes straight lines $[k_X, k_Y]$ in transverse k-space, for each reciprocal lattice vector **G**. In CBED, the incidence electron spot covers a circular region in the diffraction pattern, and, with the detector plane perpendicular to the optical axis, the allowed scattering conditions are visible as deficit lines with a distance to the origin of $r_{\text{deficit}} = (G^2/2 - k_0 G_Z)/\sqrt{G_X^2 + G_Y^2}$ and an inclination angle of $\tan(\varphi) = G_X/G_Y$.

The electrons are scattered into excess lines, which are displaced from the corresponding deficit line by the projected radial scattering vectors $[G_X, G_Y]$, so that their radial distance becomes $r_{\text{excess}} = r_{\text{deficit}} + \sqrt{G_X^2 + G_Y^2}$.

Considering rotation angles $[\alpha, \beta, \gamma] = [1.46°, 8.05°, 22.9°]$ of the graphite crystal and an initial convergence angle of 25 mrad (half angle), the precise positions (radius and inclination) of deficit and excess Bragg lines in the diffractograms can be reproduced, allowing for an assignment of the indices $h$, $k$ and $l$ (cf. Figure 1c,d).

A time-dependent deformation of the unit cell can be described by applying the distortion tensor $F(\Delta t)$ to the undistorted real-space basis of the graphite lattice $\mathbf{B}_t(\Delta t) = F(\Delta t) \cdot \mathbf{B}_u$. Generally, the average unit cell deformation within the electron beam probing volume (cf. Figure 3 and 4), can be extracted by applying a forward least squares regression analysis, fitting the absolute change in Bragg line positions $\Delta r_{(hkl)} = r_{\text{excess},(hkl),\text{exp}} - r_{\text{excess},(hkl),\text{calc}}(F)$ and inclination angles $\Delta\varphi_{(hkl)} = \varphi_{(hkl),\text{exp}} - \varphi_{(hkl),\text{calc}}(F)$ of the most intense lines with the components of the distortion tensor F as free parameters

$$\sum_{[h,k,l]} \left(\Delta r_{(hkl)}(\mathrm{F})\right)^2 + \sum_{[h,k,l]} \left(\Delta \varphi_{(hkl)}(\mathrm{F})\right)^2 \to \min .$$

No change of the inclination angles $\varphi_{(hkl),\exp}$ is observed, and we therefore adopt $\Delta\varphi_{(hkl)} \mathrel{!}= 0$ in the fitting procedure.

In our experiment, the radial position of the excess Bragg lines exhibits a high sensitivity to changes of the displacement field **u** along the Z direction, mainly related to the components $F_{xz} = (\varepsilon+\omega)_{xz}$, $F_{yz} = (\varepsilon+\omega)_{yz}$ and $F_{zz} = 1 + \varepsilon_{zz}$ of the distortion tensor **F**. Adapting these tensor components, we can quantitatively reproduce the center-of-mass shift of the selected excess lines (cf. reconstructed line positions in Figure 3c).

We note, that a pure membrane shear along the *x*-axis is described by a finite value of $F_{xz}$ and $F_{zx} = 0$, so that $\varepsilon_{xz} = \omega_{xz}$.

Generally, in electron diffraction, rocking curves can be strongly affected by multiple scattering processes. However, for the strain tensor analysis, we only evaluated Bragg lines with extinction lengths $\xi > 420$ nm, which is significantly larger than the membrane thickness (cf. Table S1), so that multiple scattering effects are minimized.

| (*hkl*) | $|\mathbf{G}_{hkl}|$ (1/nm) | $G_{ip}$ (1/nm) | $G_{op}$ (1/nm) | $\xi$ (nm) |
|---|---|---|---|---|
| $(\bar{3}\bar{2}\bar{1})$ | 12.5020 | 12.4128 | 1.4908 | 688 |
| $(\bar{3}\bar{1}\bar{1})$ | 12.5020 | 12.4128 | 1.4908 | 687 |
| $(\bar{2}\bar{4}0)$ | 16.2522 | 16.2522 | 0 | 488 |
| $(\bar{1}30)$ | 12.4128 | 12.4128 | 0 | 1177 |
| (020) | 9.3832 | 9.3832 | 0 | 716 |
| $(1\bar{3}1)$ | 16.9814 | 16.9159 | 1.4908 | 1226 |
| (201) | 9.5009 | 9.3832 | 1.4908 | 423 |
| (241) | 16.3205 | 16.2522 | 1.4908 | (kinematically forbidden) |
| (422) | 16.5235 | 16.2522 | 2.9816 | 503 |
| (452) | 21.7054 | 21.4997 | 2.9816 | 833 |

**Table S1: List of evaluated Bragg lines.** Miller indices (*hkl*) of respective reciprocal lattice plane, absolute value, in- and out-of-plane components of scattering vector $|\mathbf{G}_{hkl}|$ (1/nm notation) and extinction length $\xi$ (calculated for 120 kV electron kinetic energy).

**SI 5: Debye Waller analysis**

In order to extract the optically induced temperature rise of the graphite membrane, we analyzed the decrease of the integrated diffraction intensity $I$ within the (452) Bragg line (Figure 3b). At late delay times, a quasi-equilibrium phonon temperature is established, resulting in a nearly constant intensity drop of 41% at the edge and 25% in the continuous film, respectively (Figure S4a,b). Membrane cooling occurs on nanosecond to microsecond time scales, and is not observed in the temporal window considered here.

For thermalized phonon distributions, the relative intensity decrease $I(T_0 + \Delta T)/I(T_0)$ at temperatures $T_0$ and $T_0 + \Delta T$ is given by $\exp(2(W(T_0) - W(T)))$, in which $W(T)$ is the Debye-Waller factor. In the Debye model $W(T)$ can be expressed as

$$W(T) = \frac{\langle u_1^2 \rangle G^2}{4} = \frac{3\hbar^2 G^2}{2mk_B \Theta_D}\left[\frac{1}{4} + \left(\frac{T}{\theta_D}\right)^2 \int_0^{\theta_D/T} \frac{s}{\exp(s)-1}ds\right],$$

where $\langle u_1^2 \rangle$ is the atomic means square displacement, $m$ is the atomic mass of carbon, $k_B$ the Boltzmann constant, $\hbar$ the reduced Planck constant and $\theta_D$ the Debye temperature. For the chosen Bragg condition, the intensity decrease is dominated by thermal in-plane vibrations ( $\Theta_D \approx \Theta_{D,\,in\text{-}plane} = 1300\,K$ ),[9] resulting in an approximately linear temperature dependence for the relevant temperature range (Figure S4c). The total drop of diffracted intensity at later delay times ($\Delta t > 100$ ps) corresponds to a laser induced increase of the lattice temperature from $T_0$ = 300 K to about $\Delta T_{edge}$ = 480 K at the edge and $\Delta T_{cont}$ = 270 K in the continuous film, respectively.

As a cross check, we estimate the optically induced temperature increase from the known material constants and the incident pump fluence. Specifically, for an optical reflectivity of 25% (p-polarized, 47° incidence angle), we obtain an absorbed fluence of 8.4 mJ/cm², resulting in a temperature increase in the continuous part of the membrane $\Delta T_{cont}$ of 225 K – 450 K, depending on the base temperature of the membrane (300 K – 800 K). We note that at the employed laser repetition rate, an increase of the membrane base temperature is expected, depending on the thermal contact between the graphite flake and the supporting copper grid.

In our experimental data, we can directly quantify the increase in atomic mean square displacement in the in-plane direction, by extracting the temporal change of overall diffraction intensity into a specific Bragg line, as shown in Figure 3b. For early delay times, a biexponential decay behavior with time constants of $\tau_1$ = 850 (800) fs and $\tau_2$ = 14.6 (14.9) ps is observed at the membrane edge (continuous membrane). The extracted time scale of the fast component represents an upper bound, due to the convolution of the intrinsic dynamics with the 700-ps electron pulse duration (FWHM). This biexponential behavior was previously attributed to the thermalization of a non-equilibrium phonon distribution[10–12] generated by the ultrafast decay of strongly coupled optical phonons (SCOPs).[10–15]

Considering the out-of-plane thermal expansion coefficient of graphite $\alpha = 29 \cdot 10^{-6}$ 1/K[16], the observed change in interlayer spacing corresponds to a temperature increase of ~275 K in the continuous part of the membrane and ~450 K close to its edge, in agreement with the temperatures extracted from the Debye-Waller analysis.

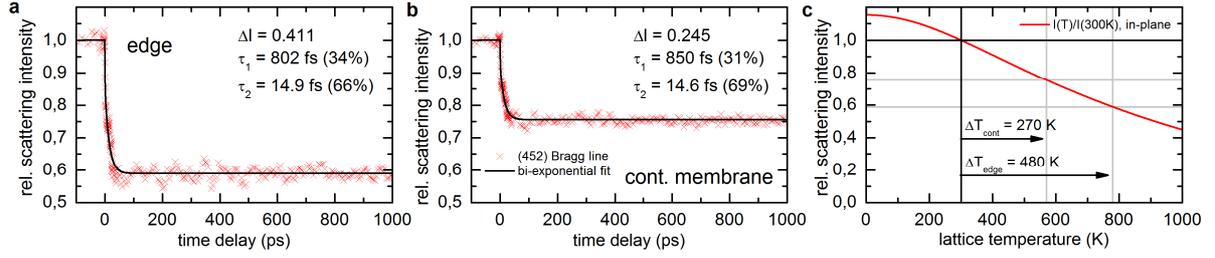

**Figure S4: Intensity change of the (452) Bragg line.** After ultrashort excitation with an illuminating fluence of 16.2 mJ/cm², the scattering efficiency of the 452 Bragg line decreases in a bi-exponential behavior both at the edge (a) and in the continuous film (b). Time constants $\tau_1$ and $\tau_2$ obtained from bi-exponential fit. Ratio of fast/slow component relative to the overall intensity drop $\Delta I$ given in parenthesis. The different magnitudes of the overall drop (a) at the edge (41.1%) and (b) in the continuous film (24.5%) reflects the difference in local temperature. (c) The scattering intensity for the (452) drops with an increase of equilibrium lattice temperature after laser heating of the graphite membrane (considering the in-plane Debye temperature).

## SI 6: Numerical simulation of graphite thin film lattice dynamics

To complement our experiments, we calculate quantitative Bragg line profiles from the inhomogeneous strain distributions derived from numerical simulations (Figure 4c and 5a,b, right panel). To this end, we solve the elastodynamic wave equation in a 2D finite element simulation (COMSOL Multiphysics 5.3, simulated slab size of 117 nm × 8 µm) in response to an optically induced thermal stress $\sigma_{ij}(x,z,t) = \Delta T_{\text{eff}}(x,z,t) \cdot \sum_k C_{ijkk} \alpha_{kk}$,[17] considering graphite bulk properties for the elasticity tensor **C**,[18] and the diagonal components of the thermal expansion tensor $\boldsymbol{\alpha}_{xx,yy,zz} = [1, 29, 29] \cdot 10^{-6}\ \text{K}^{-1}$.[16]

The temperature field is obtained by taking into account the inhomogeneously deposited optical excitation and the graphite heat capacity.[19] In addition, diffusional heat transport was included in the model using an anisotropic heat conductivity.[19]

From the resulting time- and position-dependent, three-dimensional displacement field **u**(*x, z, t*), we calculate the diffraction rocking curve of a specific Bragg scattering condition (*hkl*), according to Eq. 1 in the main text (cf. Ref. 20,21).

In a first model for the temperature field $\Delta T_{\text{eff}}(x, z, t)$, we utilized the depth-dependent absorbed optical power $Q(z, t)$ in the membrane, including the optical absorption length of graphite ($\delta_P = 36$ nm)[22] at the employed wavelength and incidence angle, and Fabry-Perot interference at the membrane faces. To phenomenologically model the coupling time between initial optical excitation and out-of-plane expansional motion, we assumed $Q \propto \exp(-t/\tau)$, with $\tau = 8$ ps. While this model generally reproduces the overall Bragg line displacements (observed in the continuous membrane) as well as the asymmetry at the crests of the Bragg line oscillations, the line splitting is largely overestimated. A good match with the experimental profiles is obtained, when spreading the optically excitation $Q$ to about 90 nm within the depth of the film (Figure 5a). Possible mechanisms include the fast ballistic transfer of heat, in qualitative agreement with theoretical predictions[23] and experimental findings[24] for the thermal phonon mean-free-path at room temperature along the out-of-plane direction in the 100-nm range. We note, that such a process also contains a temporal component governed by the group velocity of thermal phonons in the out-of-plane direction, which warrants further studies.

In addition, using a thermal stress model, a good agreement with the experimental line splittings at the membrane edge is also found (Figure 5b). To account for the lateral inhomogeneous optical excitation field, we scaled the strain tensor components to match the experimental amplitude of the expansional and shear mode at the edge. A better fit to the experimental profiles is found, if adopting a shorter optical excitation depth of 36 nm and a coupling time $\tau = 1$ ps for the in-plane component of the displacement field, suggesting a strong mode-specificity of the Grüneisen parameter[25] for the early non-equilibrium phonon distribution.

**SI 7: Characterization of the optical near-field structure at the graphite edge**

In order to connect the experimentally observed, laterally inhomogeneous strain $\varepsilon_{zz}$ to the optical near-field at edge, we characterize the local optical field distribution at time zero (Figure 4d) by a scanning variant[26,27] of photon induced near-field electron microscopy (PINEM).[27–31] This technique gives access to a specific spatial Fourier component of the optical field component $E_z$ along the propagation direction of the electron beam, which is quantified by a coupling constant $|g|$, serving as a measure for the nanoscale structure of the scattered light field at the sample. Specifically, with focused electron probe pulses, we record electron energy spectrum at a varying distance from the membrane edge (cf. Figure S5a). Under optical illumination (800-nm central wavelength, dispersively stretched 3.4-ps optical pulse duration) and at $\Delta t = 0$, the spectra exhibit multiple sidebands spaced by the photon energy with the population of sidebands related to $|g|$.[27,30,31]

As shown in Figure 4d, we find the coupling constant to vary with the distance to the graphite edge in the form of a damped oscillation with a 2-µm spatial periodicity. Generally, optical scattering from the edge of thin film is a complex problem, but the spatial periodicity of the light field at the surface can be obtained by considering the interference of the incident illumination wave vector $\mathbf{k}_0 = 2\pi/\lambda$ with the light field scattered at the edge with wavevector $\mathbf{k}_1$. The interfering light field exhibits a beating in its amplitude along the interface with a periodicity $d$ given by the difference wave vector $|\Delta\mathbf{k}| = 2\pi/d = |\mathbf{k}_{0,\|} - \mathbf{k}_1|$, where $\mathbf{k}_{0,\|}$ is the projected incident wave vector onto the interface. With $|\mathbf{k}_1| = |\mathbf{k}_0|$ and $|\mathbf{k}_{0,\|}| = |\mathbf{k}_0| \cdot \sin(\alpha)$, the periodicity $d = 1/(1-\sin(\alpha))$ is obtained. For the experimental incidence angle alpha of about 39°, a spatial periodicity $d$ of 2.16 µm is expected in reasonable agreement with the scanning PINEM maps. For the ultrafast scanning CBED experiments (Figure 4a,b,e), an optical incidence angle of 47° was used, resulting in a modulated optical field strength with a periodicity of about 3 µm, closely reproducing the observed spatially modulation in the expansional mode amplitude (Figure 3b).

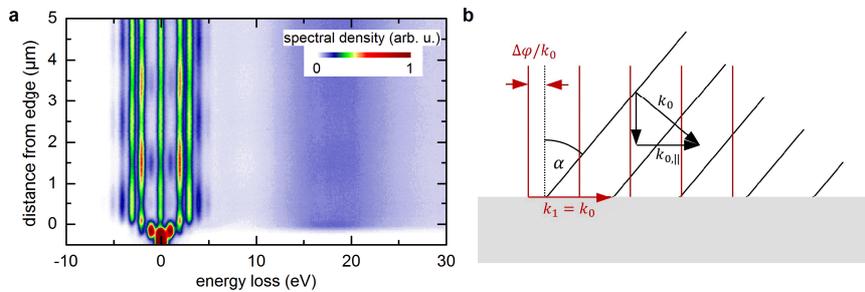

**Figure S5: Characterization optical near-field at the graphite edge.** (a) Normalized electron energy loss spectra (EELS) at spatial and temporal overlap of electron and laser beam as a function of distance from the edge. The near-field extends into the vacuum, which is evident by regarding the scattering from graphite bulk plasmons at higher energy losses (>10 eV). (b) The variation of the optical amplitude origins from the interference of the impinging light field with the scattered wave from the graphite edge. The maxima reoccur for $d = 1/(1-\sin(\alpha))$.